\documentstyle[11pt,epsbox]{article}

\makeatletter

\def\lsim{\compoundrel<\over\sim}
\def\compoundrel#1\over#2{\mathpalette\compoundreL{{#1}\over{#2}}}
\def\compoundreL#1#2{\compoundREL#1#2}
\def\compoundREL#1#2\over#3{\mathrel
  {\vcenter{\hbox{$\m@th\buildrel{#1#2}\over{#1#3}$}}}}
\makeatother

\title{{\bf Constraints of mixing angles 
from neutrino oscillation experiments and neutrinoless double 
beta decay}}

\author{Takeshi Fukuyama \footnotemark[1],~~Kouichi Matsuda  
\footnotemark[1]~~ 
and Hiroyuki Nishiura \footnotemark[2]  
                   \\
        $\ast$~~Department of Physics, \\
        Ritsumeikan University, Kusatsu, \\
        Shiga, 525 Japan \\
        $\dagger$~~ Department of General Education, \\
        Junior College of Osaka Institute of Technology, \\
        Asahi-ku, Osaka,535 Japan}

\date{}

\begin{document}

\maketitle



\begin{abstract}
 From the analyses of the recent data of neutrino oscillation experiments 
(especially the CHOOZ and the Super KAMIOKANDE experiments), we discuss how 
these data affect  the neutrinoless double beta decay 
($(\beta \beta)_{0\nu}$) rate  and vice versa assuming that neutrinos are 
Majorana particles. 
For the case that $m_1 \sim m_2 \ll m_3$ ($m_i$ are neutrino masses), we 
obtain, from the data of the CHOOZ and Super KAMIOKANDE, 
$0.28 \lsim \sin^2\theta_{23} \lsim 0.76$ and $\sin^2\theta_{13} \lsim 
0.05$. Combining the latter constraint with the analysis of 
the "averaged" neutrino mass 
\(\langle m_\nu \rangle\) appeared in $(\beta \beta)_{0\nu}$,
we find that 
\(\frac{\langle m_\nu \rangle-m_2}{m_3-m_2}<\sin^2 \theta_{13} \lsim 0.05\), 
which leads to the constraint on \(\langle m_\nu \rangle\) as 
\(\langle m_\nu \rangle \lsim 0.05 m_3+(1-0.05)m_2\).
For the case that $m_1 \ll m_2 \sim m_3$, 
we find that the data of neutrino oscillation experiments and 
$(\beta \beta)_{0\nu}$ imply the following constraints of mixing angles.
If \(0.95 m_3 \lsim \langle m_{\nu} \rangle < m_3 \), 
\(0 \le \sin^2\theta_{13} \le \frac{m_3+\langle m_{\nu} \rangle}{2m_3}\). 
If \(\langle m_{\nu} \rangle \lsim 0.95 m_3\), 
\(\frac{\langle m_{\nu} \rangle-(1-0.05)m_3}
 {\langle m_{\nu} \rangle-(1+0.05)m_3}
\lsim \sin^2\theta_{12} \le 1 \) and
\(\frac{(1-0.05)m_3-\langle m_{\nu} \rangle}{2m_3}
\lsim \sin^2\theta_{13} \le \frac{m_3+\langle m_{\nu} \rangle}{2m_3} \).
\\
PACS number(s): 14.60Pg 11.30.Er 13.35.Hb 23.40.Bw
\end{abstract}


\newpage



In the last decade important data of neutrino oscillation experiments 
have 
appeared.  They have come from solar neutrino oscillation\cite{kamioka} 
\cite{homestake} \cite{gallex} \cite{sage}, atmospheric neutrino deficit 
\cite{skamioka} and the neutrino oscillation from reactors 
\cite{chooz} \cite{bugey} and 
accelerators \cite{chorus} \cite{nomad}.  
Another important data come from the neutrinoless double beta decay 
(\((\beta\beta)_{0\nu}\)) \cite{particle}, 
though the neutrinos are need to be Majorana particles for this process to
occur. In the previous paper \cite{fuku}, 
we have obtained the constraints from \((\beta\beta)_{0\nu}\) 
on the lepton mixing angles by taking possible leptonic CP violating phases into account. In this note we combine the constraints from \((\beta\beta)_{0\nu}\) with those from the recent CHOOZ reactor experiment \cite{chooz}, 
the Super KAMIOKANDE atmospheric neutrino experiment \cite{skamioka} and the solar neutrino experiment 
\cite{fogli}.
Before proceeding to this arguments we briefly review the formulation 
developed 
in \cite{fuku} which is necessary for the present study.
We assume neutrinos are Majorana particles.  For Majorana neutrinos, the 
Cabibbo-Kobayashi-Maskawa (CKM) left-handed lepton mixing matrix 
takes the following form:
\begin{equation}
U=
\left(
\begin{array}{ccc}
c_1c_3&s_1c_3e^{i\beta}&s_3e^{i(\rho-\phi )}\\
(-s_1c_2-c_1s_2s_3e^{-i\phi})e^{-i\beta}&
c_1c_2-s_1s_2s_3e^{i\phi}&s_2c_3e^{i(\rho-\beta )}\\
(s_1s_2-c_1c_2s_3e^{i\phi})e^{-i\rho}&
(-c_1s_2-s_1c_2s_3e^{i\phi})e^{-i(\rho-\beta )}&c_2c_3\\
\end{array}
\right).\label{eq3_17_1}
\end{equation}
Here $c_j=\cos\theta_j$, $s_j=\sin\theta_j$ 
($\theta_1=\theta_{12},~\theta_2=\theta_{23},~\theta_3=\theta_{31}$) and
three 
CP violating phases, $\beta$, $\rho$ and $\phi$ appear for Majorana
neutrinos.  The decay ratio of \((\beta \beta)_{0\nu}\) is, 
in the absence of right-handed couplings,
proportional to the "averaged" mass
\( \langle m_{\nu} \rangle\) defined by \cite{doi}
\begin{equation} 
\langle m_{\nu} \rangle\equiv |\sum _{j=1}^{3}U_{ej}^2m_j|. \label{eq3_17_2}
\end{equation}
Substituting the expression (\ref{eq3_17_1}) into Eq. (\ref{eq3_17_2}), 
we obtain
\begin{equation}
\langle m_{\nu} \rangle^2
=(m_1c_1^2c_3^2-m_2s_1^2c_3^2\cos2\beta'-m_3s_3^2\cos2\rho ')^2
+(m_2s_1^2c_3^2\sin2\beta '+m_3s_3^2\sin2\rho ')^2,\label{eq3_24_3}
\end{equation}    
where we have  
\begin{equation}
\beta '\equiv {\pi\over 2}-\beta ,\quad \rho '
 \equiv {\pi\over 2}-(\rho-\phi ).
\end{equation}
Rewriting $\cos2\rho '$ and $\sin2\rho '$ by $\tan \rho '$, we can 
consider
Eq. (\ref{eq3_24_3}) 
as an equation of $\tan \rho '$,
\begin{equation}
a_{+\beta}\tan^2\rho '+b_{\beta}\tan \rho '+a_{-\beta}=0.
\label{eq3_24_5}
\end{equation}  
Here $a_{\pm \beta}$ and $b_{\beta}$ are defined by
\begin{eqnarray}
a_{\pm \beta}&\equiv& 4\sin^2\beta 'm_2s_1^2c_3^2(m_1c_1^2c_3^2\pm 
m_3s_3^2)+(m_1c_1^2c_3^2-m_2s_1^2c_3^2\pm m_3s_3^2)^2-\langle m_{\nu} 
\rangle^2\\
b_\beta&\equiv& 4m_2m_3s_1^2s_3^2c_3^2\sin2\beta '.\nonumber
\end{eqnarray}
So the discriminant $D$ for Eq. (\ref{eq3_24_5})
must 
satisfy the following inequality:
\begin{eqnarray}
D&\equiv& b_\beta^2-4a_{+\beta}a_{-\beta}\nonumber\\
&=&4^3(m_1c_1^2c_3^2)^2(m_2s_1^2c_3^2)^2(f_+-\sin^2\beta ')(\sin^2\beta
'-f_-)\geq 
0,
\end{eqnarray}
where 
\begin{equation}
f_{\pm}\equiv {(\langle m_{\nu} \rangle\pm 
m_3s_3^2)^2-(m_1c_1^2c_3^2-m_2s_1^2c_3^2)^2\over
4m_1m_2c_1^2s_1^2c_3^4}
\end{equation}
So we obtain
\begin{equation}
f_-\leq \sin^2\beta '\leq f_+.
\end{equation}
It follows from Eq. (9) that
\begin{equation}
f_-\leq 1,~~ f_+\geq 0.\label{eq3_17_10}
\end{equation}
Analogously, rewriting $\cos2\beta '$ and $\sin2\beta '$ as $\tan \beta
'$, we 
obtain the inequalities:
\begin{equation}
g_-\leq 1,~~ g_+\geq 0.\label{eq3_17_11}
\end{equation}
Here
\begin{equation}
g_{\pm}\equiv {(\langle m_{\nu} \rangle\pm 
m_2s_1^2c_3^2)^2-(m_1c_1^2c_3^2-m_3s_3^2)^2\over
4m_1m_3c_1^2s_3^2c_3^2}.
\end{equation}
Using the inequalities (10) and (11), we can determine the allowed 
region for 
mixing 
angles in the $s_1^2$ versus $s_3^2$ plane once the neutrino masses $m_i$  
and 
the averaged neutrino mass \(\langle m_{\nu} \rangle\) are known.

Without loss of generality, neutrino masses are ordered as \(m_1 \le m_2 
\le 
m_3\), and $\langle m_\nu \rangle$ belongs to one of the following three 
cases:

(a) $\langle m_\nu \rangle \leq m_1$,
\par

(b) $m_1 \leq \langle m_\nu \rangle \leq m_2$
\par
and
\par
(c) $m_2 \leq \langle m_\nu \rangle \leq m_3$.

Note that the definition of $\langle m_{\nu}\rangle $ in Eq. (\ref{eq3_17_2}) 
and the Schwartz inequality jointly imply that
\begin{equation}
\langle m_\nu \rangle \leq \sum _{j=1}^{3} m_j|U_{ej}^2|\leq m_3\sum
_{j=1}^{3} 
|U_{ej}^2|=m_3, \label{eq16}
\end{equation}
that is, $\langle m_\nu \rangle$ can not be larger than $m_3$.
The allowed regions in the 
$s_1^2$ versus $s_3^2$ plane 
for cases (a), (b) and (c) are obtained from 
Eqs. (\ref{eq3_17_10}) and (\ref{eq3_17_11}), and are shown in Fig. 1.
From Fig. 1, we obtain the following bounds on $s_3^2$ :
\begin{eqnarray}
\frac{\langle m_{\nu} \rangle-m_2}{m_3-m_2} 
\leq s_3^2\leq \frac{m_2+\langle m_{\nu} \rangle}{m_3+m_2} \quad
&\mbox{for}& \quad m_2 \le \langle m_\nu \rangle
\label{eq17}\\
0 \leq s_3^2\leq \frac{m_2+\langle m_{\nu} \rangle}{m_3+m_2} \quad
&\mbox{for}& \quad \langle m_\nu \rangle \le m_2 
\nonumber
\end{eqnarray}

So far we have presented a brief review of the constraints from 
\((\beta\beta)_{0\nu}\). 
This method is 
very general and we have not imposed any concrete data on it. 
It is worthwhile to note that the allowed region in the 
\(s_3^2\) versus \(\langle m_\nu \rangle\) plane is obtained from 
Eq. (\ref{eq17}) and is shown in Fig. 2.
Next we proceed to 
discuss how the recent data from 
the neutrino oscillation experiments yield additional 
restrictions on the 
mixing angles. 
We consider the data of the CHOOZ reactor experiment \cite{chooz} and 
the Super KAMIOKANDE atmospheric neutrino experiment \cite{skamioka}. 
The latter indicates 
$\delta m^2=0.01 \sim 0.001$ 
eV$^2$ and the former gives severe restriction on the mixing angle in 
this mass region
\begin{equation}
\sin^2 2\theta\lsim 0.2. \label{eq2_28_2}
\end{equation}
Hereafter, let us discuss following two scenarios for the mass hierarchy,
\par
(A)$\quad m_1\sim m_2\ll m_3$,
\par
(B)$\quad m_1\ll m_2\sim m_3$.
\par
First we argue case (A).
In three generation model,
 \(P(\bar{\nu}_e\rightarrow \bar{\nu}_e)\) is given 
by
\begin{equation}
P(\bar{\nu}_e\rightarrow \bar{\nu}_e)
 =1-4s_3^2c_3^2\sin^2\Bigl( {\delta m_{13}^2L \over 4E}\Bigr),
\end{equation}
for case (A). So $\theta=\theta_3$ in Eq. (15) and
\begin{equation}
0 \le s_3^2 \lsim 0.05
 \quad \mbox{or}
 \quad 0.95 \lsim s_3^2  \le 1.
\end{equation}

On the other hand, Super KAMIOKANDE indicates in
\begin{equation}
P(\nu_{\mu}\rightarrow \nu_{\mu})
 =1-4c_3^2s_2^2(1-c_3^2s_2^2)\sin^2\Bigl( {\delta m_{13}^2L \over 
4E}\Bigr),
\end{equation}
that \(0.8 \lsim 4 c_3^2 s_2^2(1-c_3^2 s_2^2) \le 1\), namely
\begin{equation}
{0.28 \over 1-s_3^2} \lsim s_2^2 \lsim {0.72 \over 1-s_3^2}.
\end{equation}
Then, from these two experiments we obtain
\begin{equation}
0.28 \lsim s_2^2 \lsim 0.76,\quad s_3^2 \lsim 0.05.\label{eq3_17_020}
\end{equation}
It should be noted that the data of solar neutrino impose no further constraint \cite{fogli}. These results are depicted in Fig. 3.
Now combining the allowed region in Fig. 3 with one in Fig. 2 from 
\((\beta\beta)_{0\nu}\), we obtain a further constraint.
It follows from Eqs. (\ref{eq17}) and (\ref{eq3_17_020}) that 
\begin{equation}
\frac{\langle m_\nu \rangle - m_2}{m_3-m_2} \le s_3^2 \lsim 0.05.
\end{equation}
Namely, we find 
\begin{equation}
\langle m_{\nu} \rangle \lsim 0.05m_3+(1-0.05)m_2 \simeq 0.05 m_3.
\label{eq2_24_22}
\end{equation}
Since the Super KAMIOKANDE experiment indicates 
\(\delta m^2 \simeq m_3^2 = 0.01 \sim 0.001 \mbox{eV}^2\), 
we obtain from Eq. (\ref{eq2_24_22}) that 
\(\langle m_\nu \rangle < 0.005 \sim 0.0005 \mbox{eV}\) 
in case (A).

Next let us consider case (B). 
The vacuum oscillation probability 
\(P(\nu_\alpha \to \nu_\alpha)\) of case (B) 
is obtained from that of case 
(A) by the permutation of \(|U_{e3}| \leftrightarrow |U_{e1}|\).
Thus  \(P(\bar{\nu}_e\rightarrow \bar{\nu}_e)\) for case (B) is given 
by
\begin{equation}
P(\bar{\nu}_e\rightarrow \bar{\nu}_e)
 =1-4c_1^2c_3^2(1-c_1^2c_3^2)\sin^2\Bigl( {\delta m_{13}^2L \over 
4E}\Bigr).
\end{equation}
The CHOOZ experiment gives the constraint that 
\(4c_1^2c_3^2(1-c_1^2c_3^2) \lsim 0.2\). Namely, we find 
\begin{equation}
1-{0.05\over 1-s_1^2} \lsim s_3^2
\end{equation}
or
\begin{equation}
s_3^2 \lsim 1-{0.95\over 1-s_1^2}. \label{eq3_34_25}
\end{equation}
On the other hand, the Super KAMIOKANDE experiment indicates in  
\begin{equation}
P(\nu_{\mu}\rightarrow \nu_{\mu})
 =1-4|U_{\mu 1}|^2(1-|U_{\mu 1}|^2)\sin^2\Bigl( {\delta m_{13}^2L \over 
4E}\Bigr)
\end{equation}
that \(0.8 \lsim 4|U_{\mu 1}|^2(1-|U_{\mu 1}|^2) \le 1\), namely
\begin{equation}
0.28 \lsim |U_{\mu 1}|^2 \lsim 0.72. \label{eq3_17_27}
\end{equation}
Here $|U_{\mu 1}|^2$ is 
\begin{equation}
|U_{\mu 1}|^2=s_1^2c_2^2+c_1^2s_2^2s_3^2+2s_1s_2s_3c_1c_2\cos \phi
\end{equation}
from Eq. (1). Hence we obtain
\begin{equation}
(s_1c_2-c_1s_2s_3)^2<|U_{\mu 1}|^2<(s_1c_2+c_1s_2s_3)^2. \label{eq3_17_29}
\end{equation}
Then from Eqs. (\ref{eq3_17_27}) and (\ref{eq3_17_29}) it follows that
\begin{equation}
(s_1c_2-c_1s_2s_3)^2 \lsim 0.72
\end{equation}
and
\begin{equation}
0.28 \lsim (s_1c_2+c_1s_2s_3)^2.  \label{eq3_17_31}
\end{equation}
Since $(s_1c_2+c_1s_2s_3)^2<s_1^2+c_1^2s_3^2$ we obtain from 
Eq. (\ref{eq3_17_31})
\begin{equation}
{0.28-s_1^{2}\over 1-s_1^{2}} \lsim s_3^2. \label{eq3_17_32}
\end{equation}
The constraint Eq. (\ref{eq3_17_32}) 
from the Super KAMIOKANDE experiment excludes 
the small angle region, Eq. (\ref{eq3_34_25}), of the allowed regions 
from the CHOOZ experiment. 

As for solar neutrino oscillation, Fogli et al. \cite{fogli} discussed only case (A). 
Arguments for case (B) needs new assumption and here we do not take the 
data of solar neutrino experiments for case (B) into account.
Thus the constraints from the oscillation experiments are summarized in Fig. 4.
We know that the allowed region 
is restricted to the upper part of Fig. 4.
Now, let us superimpose this result on the allowed region of 
$(\beta\beta)_{0\nu}$, Fig. 1. 
In the following discussions, we restrict ourself to the case where 
\(m_1 \ll \langle m_\nu \rangle \le m_2 \sim m_3\), expecting that 
\(\langle m_\nu \rangle\) is not too small. In this case, 
the allowed region from \((\beta\beta)_{0\nu}\) 
in the \(s_1^2\) versus \(s_3^2\) plane is obtained from Fig. 1(b).
The allowed region depends on the value of \(\langle m_\nu \rangle/m_3\) 
as shown in Fig. 5.

Combining the allowed region in Fig. 5 with one in Fig. 4, 
we obtain the following allowed regions of \(s_1^2\) and of \(s_3^2\) 
in terms of \(\langle m_\nu \rangle/m_3\), which are shown in Fig. 6.
\begin{eqnarray}
&&\frac{\frac{\langle m_\nu \rangle}{m_3} -(1-0.05)}
	{\frac{\langle m_\nu \rangle}{m_3} -(1+0.05)} \lsim s_1^2 \le 1,
\quad
\frac{1}{2} \Bigl\{ (1-0.05) - \frac{\langle m_\nu \rangle}{m_3} \Bigr\}
 \lsim s_3^2 \le \Bigl(1+ \frac{\langle m_\nu \rangle}{m_3} \Bigr)
  \nonumber\\
&&\hspace{8.5cm} 
\quad \mbox{for} \ \ \frac{\langle m_\nu \rangle}{m_3} \lsim 0.95. 
\label{eq3_24_35} \\
&&\qquad 0 \le s_1^2 \le 1, \quad 0 \le s_3^2 \le 
\frac{1}{2} \Bigl( 1+ \frac{\langle m_\nu \rangle}{m_3} \Bigr) \qquad
\mbox{for} \ \ 0.95 \lsim \frac{\langle m_\nu \rangle}{m_3}.\nonumber
\end{eqnarray}
Note that since \(m_3=0.01 \sim 0.001 \mbox{eV}\) from Super KAMIOKANDE and 
\(\langle m_\nu \rangle < \mbox{O}(10^{-1}) \mbox{eV} \) from 
\((\beta\beta)_{0\nu}\) we have no constraint for 
\(\langle m_\nu \rangle/m_3\) at present.
In conclusion, the recent data of neutrino oscillation experiment, especially,
the CHOOZ and the Super KAMIOKANDE experiments are analyzed together with 
\(\langle  m_\nu \rangle\) in \((\beta\beta)_{0\nu}\) for three generations of Majorana neutrinos. We obtain an upper bound for \(\langle m_\nu \rangle\) as in Eq. (\ref{eq2_24_22}) for the case that \(m_1 \sim m_2 \ll m_3\). 
For the case that \(m_1 \ll m_2 \sim m_3\), on the other hand,
the allowed regions of \(s_1^2\) and of \(s_3^2\) in term of 
\(\langle m_\nu \rangle/m_3\) are obtained and shown in Fig. 6.

\ \\
Acknowledgements

We are grateful to O. Yasuda for calling our attention to the CHOOZ experiment.


\vspace{3cm}

\
\\
{\bf Figure Captions}\\
\ \\
{\bf Fig. 1:} The allowed region in the $s_1^2$ versus 
$s_3^2$ plane obtained from neutrinoless double beta decay is 
given by the shaded areas in the cases:
\ (a) $\langle m_\nu \rangle \leq m_1$, 
\ (b) $m_1 \leq \langle m_\nu\rangle \leq m_2 $ and 
\ (c) $m_2 \leq \langle m_\nu \rangle \leq m_3 $.

\ \\
{\bf Fig. 2:}
The allowed region in the \(s_3^2\) versus \(\langle m_\nu \rangle\) plane.    

\ \\
{\bf Fig. 3:}
The allowed regions for case (A) from the respective experiments are 
indicated by the arrows. Shaded region is the commonly accepted one. 

\ \\
{\bf Fig. 4:}
The same diagram as Fig. 3 for case (B).

\ \\
{\bf Fig. 5:}
Each allowed region (shaded region) obtained 
from neutrinoless double beta decay 
for the cases: 
\ \(\frac{\langle m_\nu \rangle}{m_3} = 0.33,\, 0.65,\, 0.98\) 
under the assumption that 
\(m_1 \ll \langle m_\nu \rangle < m_2 \sim m_3\). 
Dotted region is allowed by the CHOOZ experiment for case (B). 

\ \\
{\bf Fig. 6:} The allowed regions of \(s_1^2\) versus 
\(\langle m_\nu \rangle/m_3\) and \(s_3^2\) versus 
\(\langle m_\nu \rangle/m_3\) planes for case (B) 
obtained from Eq. (\ref{eq3_24_35}).

\clearpage

 \begin{figure}[htbp]
 	\begin{center}
 	\leavevmode
 	\epsfile{file=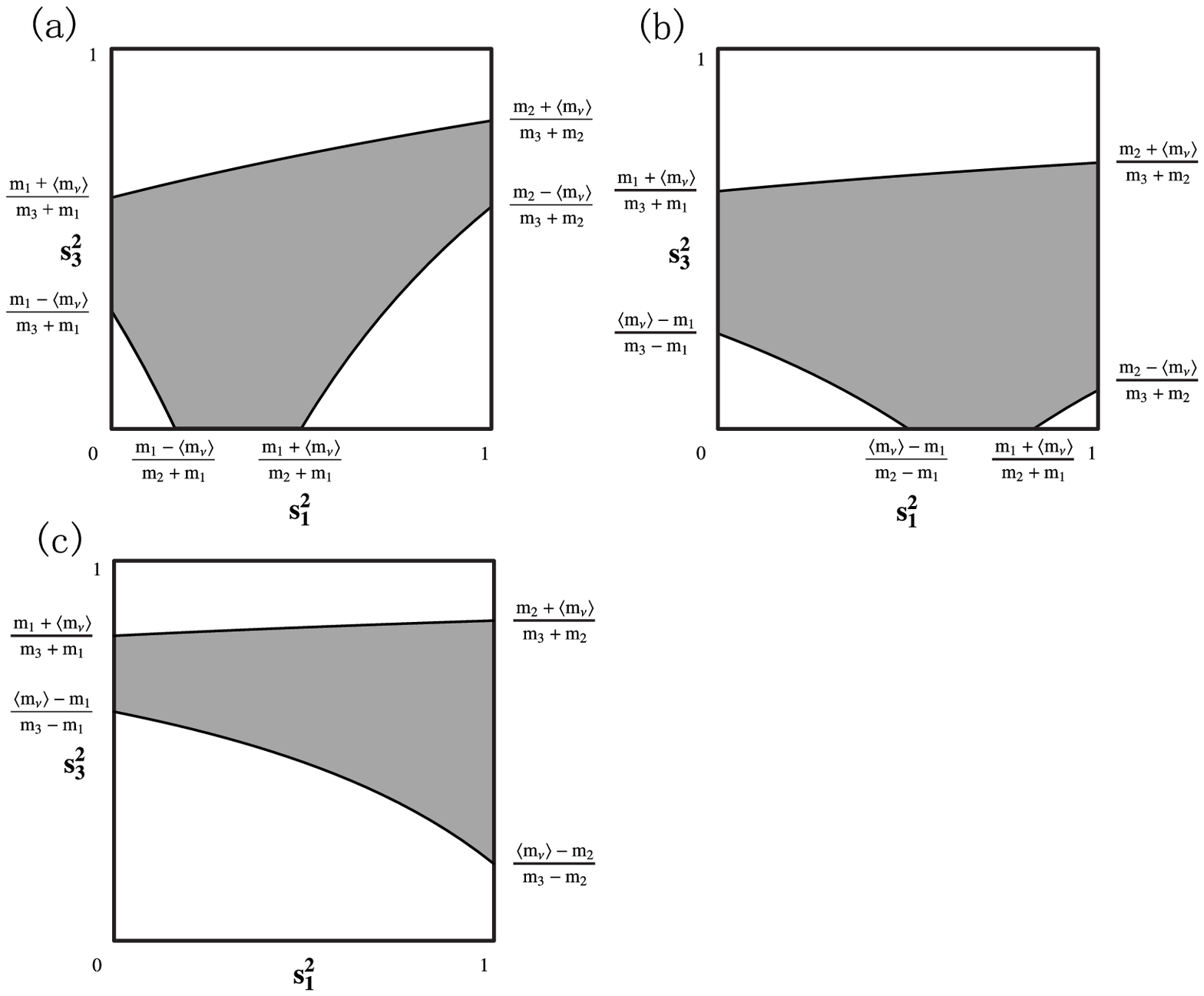,width=12cm}\\
 \ \\
 	{\Huge \bf FIG.1}
 	\end{center}
 \end{figure}

 \begin{figure}[hbtp]
 	\begin{center}
 	\leavevmode
 	\epsfile{file=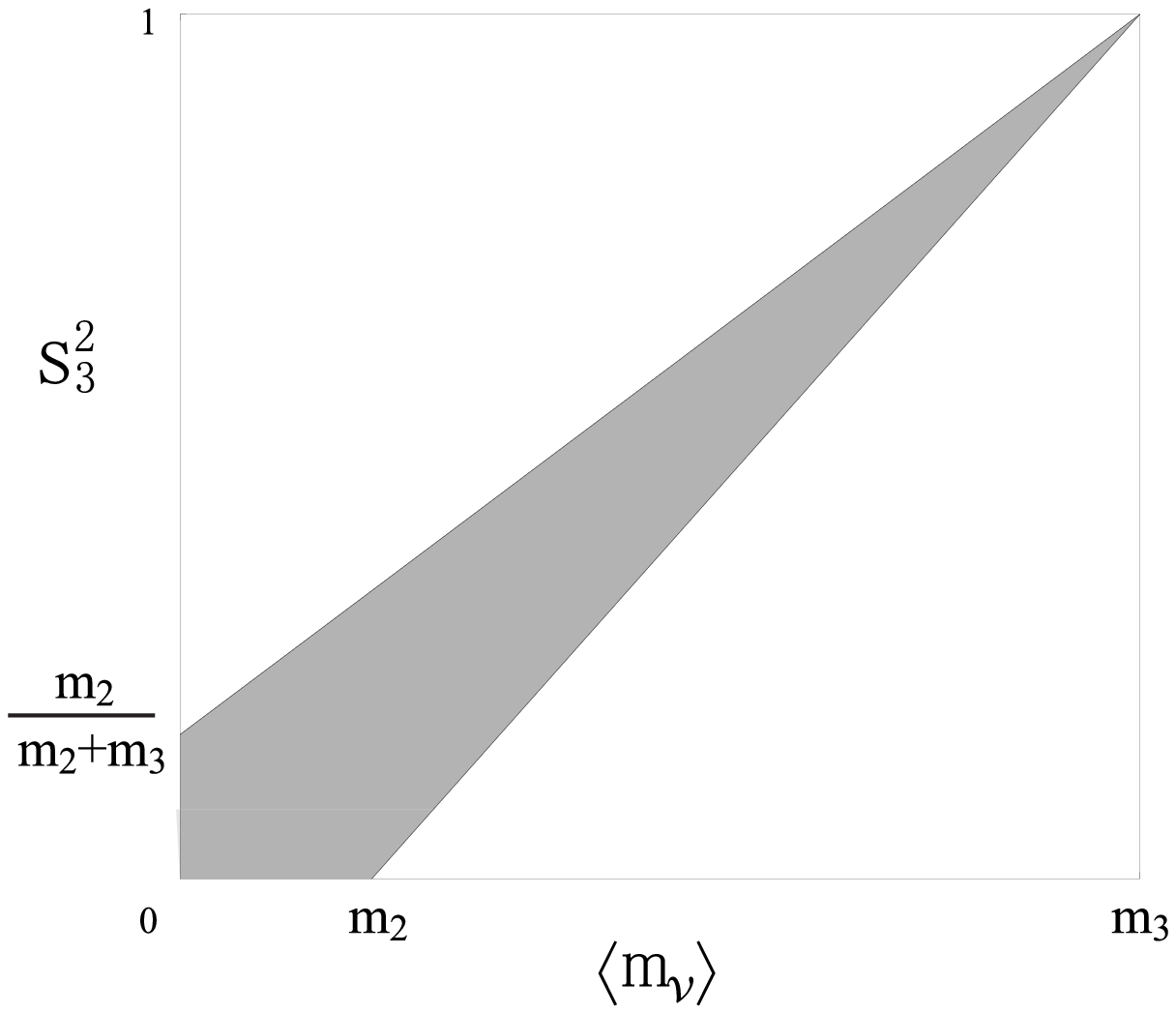,width=10cm}\\
 \ \\
 	{\Huge \bf FIG.2}
 	\end{center}
 \end{figure}

 \begin{figure}[htbp]
 	\begin{center}
 	\leavevmode
 	\epsfile{file=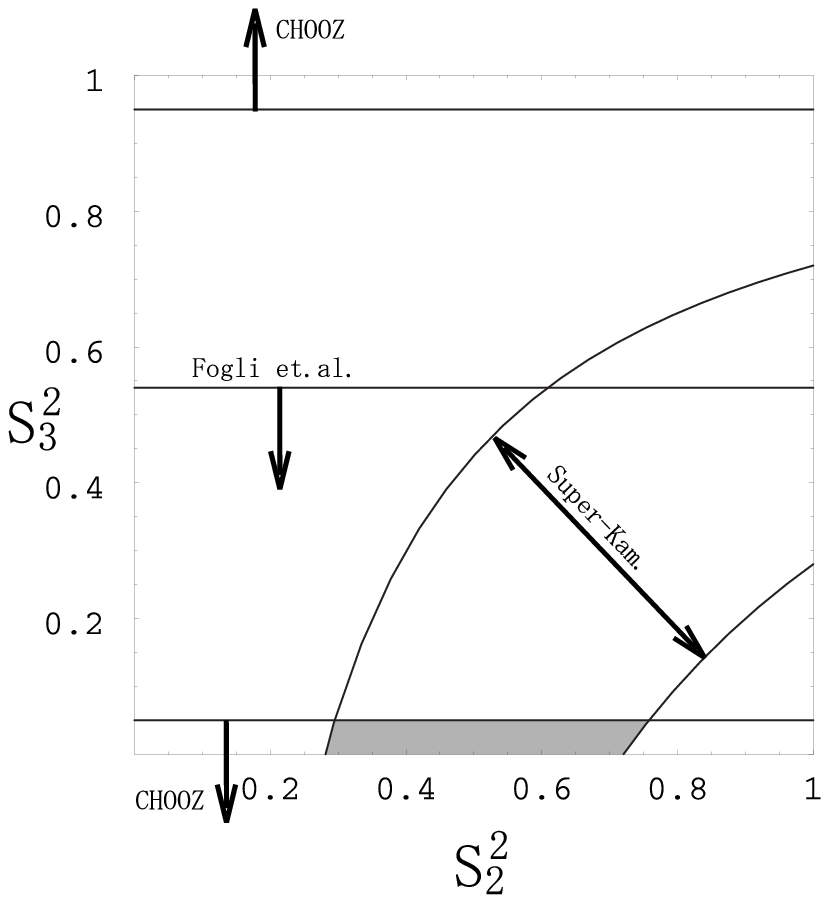}\\
 \ \\
 	{\Huge \bf FIG.3}
 	\end{center}
 \end{figure}

 \begin{figure}[htb]
 	\begin{center}
 	\leavevmode
 	\epsfile{file=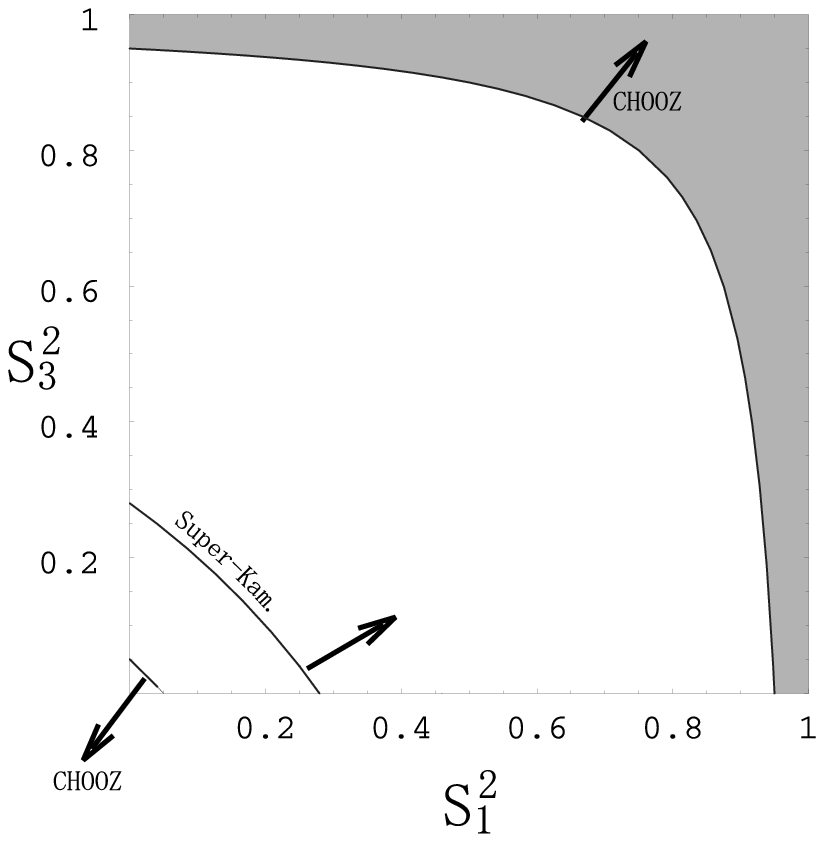}\\
 \ \\
 	{\Huge \bf FIG.4}
 	\end{center}
 \end{figure}
 \ \\
 \begin{figure}[htb]
 	\begin{center}
 	\leavevmode
 	\epsfile{file=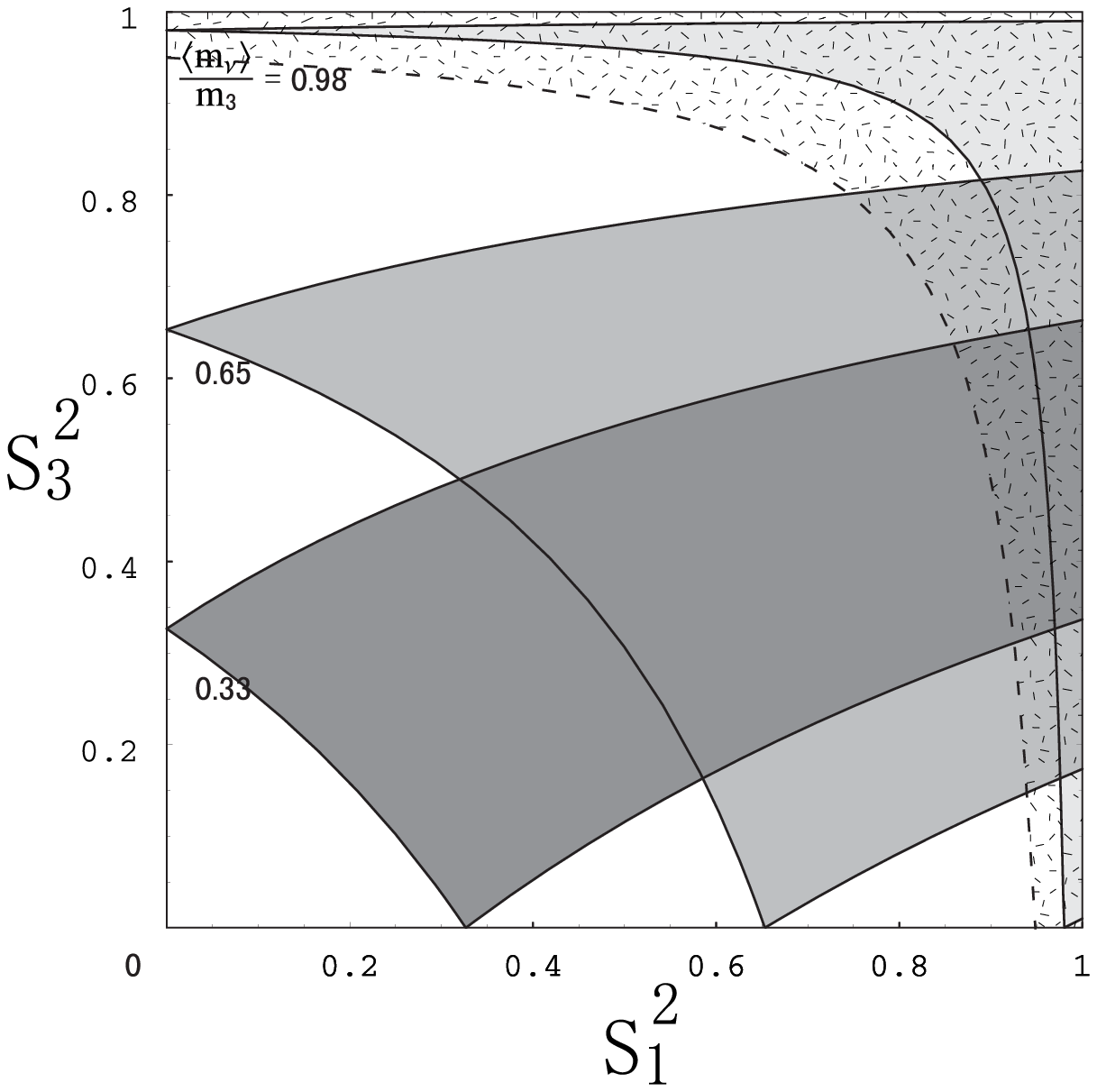,width=10cm}\\
 \ \\
 	{\Huge \bf FIG.5}
 	\end{center}
 \end{figure}

 \begin{figure}[htb]
 	\begin{center}
 	\leavevmode
 	\epsfile{file=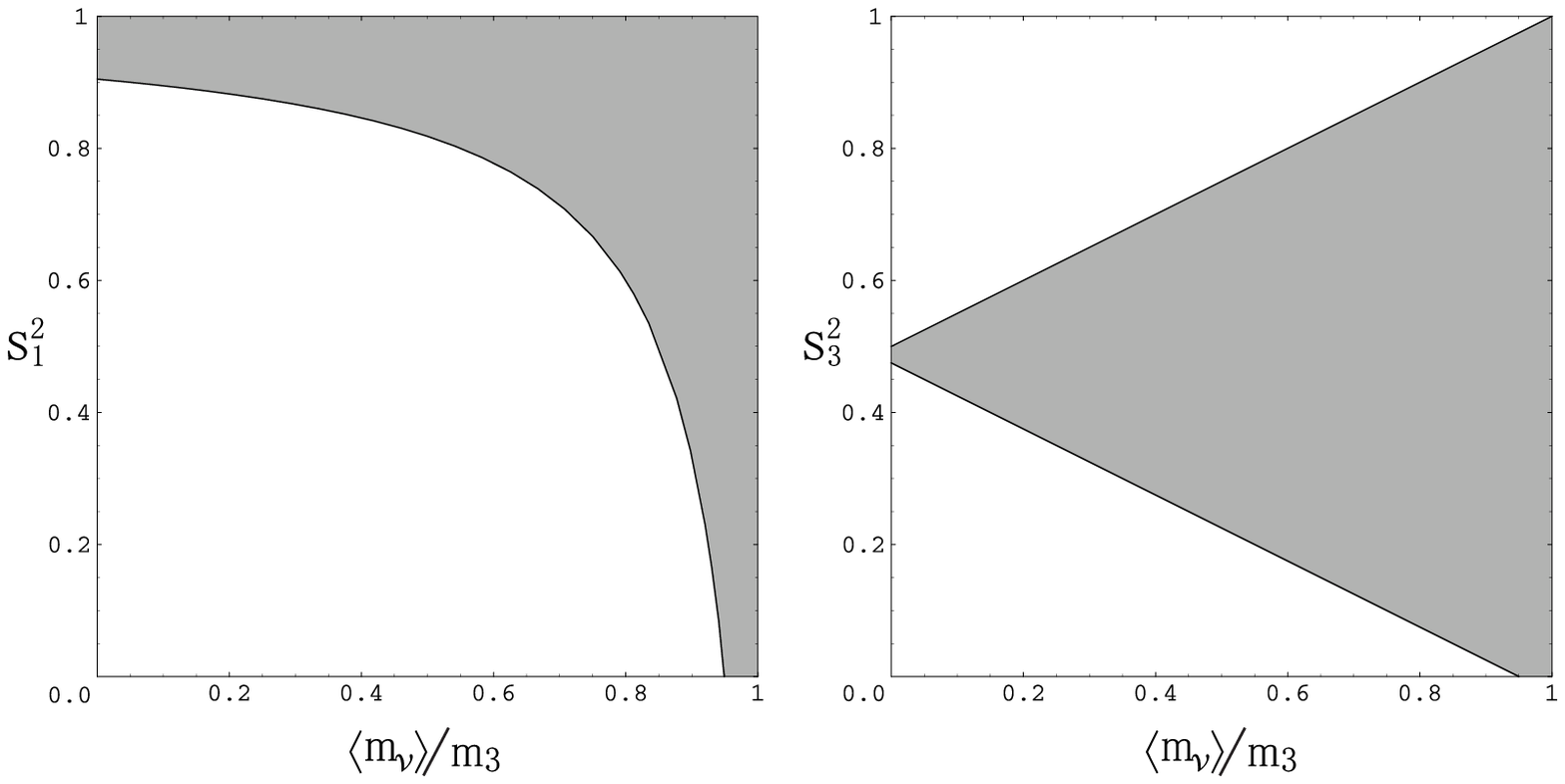,width=12cm}\\
 \ \\
 	{\Huge \bf FIG.6}
 	\end{center}
 \end{figure}


\begin{thebibliography} {99}
\bibitem{kamioka}
 K.S. Hirata et. al., Phys. Rev Lett. {\bf 63} 6 (1989); N. Hata,S. 
Bludman
and 
P. Langacker, Phys. Rev.{\bf D49} 3622 (1994); N. Hata and  P. 
Langacker,
Phys. 
Rev.{\bf D50} 632 (1994).
\bibitem{homestake}
 R. Davis Jr., Proc. of the 6'th International Workshop on Neutrino
Telescope, 
Venezia, March 1994, edited by M. Baldo Ceolin.
\bibitem{gallex}
 P. Anselmann et.al., Phys.Lett {\bf B285} 376 (1992), ibid {\bf B357} 
237
(1995).
\bibitem{sage}
 A.I. Abazov et. al., Phys. Rev. Lett. {\bf 67} 3332 (1991); J.N.
Abdurashitov 
et. al., Phys. Lett. {\bf B328} 234 (1994).
\bibitem{skamioka}
Y.Suzuki, talk at International KEK Workshop on "Kaon, Muon, Neutrino 
Physics 
and Future" Oct.31-Nov.1,1997.
\bibitem{chooz}
M.Apollonio et.al. hep-ex/9711002.
\bibitem{bugey}
B.Achkar et al.,Nucl. Phys.{\bf B434}(1995)503.
\bibitem{chorus}
 K. Winter, Nucl. Phys. {\bf B38} (Proc. Suppl.) 211 (1995).
\bibitem{nomad}
L.DiLella, Nucl. Phys.{\bf B31} (Proc. Suppl.)(1993) 319.
\bibitem{particle}
Particle Data Group, R.M.Barnet et.al., Phys. Rev. {\bf D54}, 1 (1997).
\bibitem{fuku}
T. Fukuyama, K. Matsuda and H. Nishiura, Phys. Rev. in press 
(hep-ph/9711415).
\bibitem{fogli}
G.L.Fogli,E.Lisi and D.Montanino, Phys. Rev {\bf D54} (1996) 2048.
\bibitem{doi}
M. Doi, T. Kotani, H. Nishiura, K. Okuda and E. Takasugi, Phys. Lett. 
{\bf 102B} 323 (1981); M. Doi, T. Kotani and E. Takasugi, 
Prog. Theor. Phys. Suppl. {\bf 83} 1 (1985); 
W.C. Haxton and G.J. Stophenson Jr., 
Prog. Part. Nucl. Phys. {\bf 12} 409 (1984).
\end{thebibliography}
\end{document}